\documentstyle[psfig]{mn}
\def\H0{{\it H}$_0$}
\def\Ms{{\it M}$_\odot$}

\def\q0{{\it q}$_0$}
\def\kmps{km~s$^{-1}$}

\def\ergps{erg~s$^{-1}$}
\def\kmpspMpc{km~s$^{-1}$~Mpc$^{-1}$}
\def\Ms{{\it M}$_\odot$}
\def\micron{$\mu$m}
\def\deg{$^{\circ}$}
\def\min{$^{\prime}$}
\def\sec{$^{\prime\prime}$}
\def\nH{$N_{\rm H}$\thinspace} 
\def\DelnH{$\Delta N_{\rm H}$}
\def\psqcm{cm$^{-2}$}
\def\ergpspsqcm{erg~cm$^{-2}$~s$^{-1}$}

\def\Zs{$Z_{\odot}$}

\def\cps{ct\thinspace s$^{-1}$}
\def\phpspsqcm{ph\thinspace s$^{-1}$\thinspace cm$^{-2}$}

\title[X-ray emission from Zw1718.1--0108] 
{Zw 1718.1--0108: a highly disturbed galaxy cluster at low Galactic 
latitude}
\author[K. Iwasawa et al] 
{\parbox[]{6.5in} {K. Iwasawa$^1$, S. Ettori$^1$, A.C. Fabian$^1$, 
A.C. Edge$^2$ and H. Ebeling$^3$}\\
\\
$^1$ Institute of Astronomy, Madingley Road, Cambridge CB3 0HA\\ 
$^2$ Department of Physics, University of Durham, South Road, Durham DH1 3LE\\
$^3$ Institute for Astronomy, University of Hawaii, 2680 Woodlawn Drive, Honolulu HI 96822, USA\\
}
\date{}

\begin{document}

\maketitle

\begin{abstract}
We report the discovery of highly distorted X-ray emission associated
with the nearby cluster Zw 1718.1--0108, 
one of whose dominant members is
the powerful radio galaxy 3C353. This cluster has been missed
by previous X-ray cluster surveys because of its low Galactic latitude 
($b = 19.5^{\circ}$), despite its brightness in the hard X-ray band
(2--10 keV flux of $1.2\times 10^{-11}$\ergpspsqcm).
Our optical CCD image of the central part of the cluster reveals
many member galaxies which are dimmed substantially 
by heavy Galactic extinction.
We have measured redshifts of three bright galaxies near the X-ray
emission peak and they are all found to be
around $z=0.028$. 
The ASCA GIS and ROSAT HRI images show
three aligned X-ray clumps embedded in 
low surface-brightness X-ray emission extended by 
$\sim 30$ arcmin.
The averaged temperature measured with ASCA   
is $kT=4.3\pm 0.2$ keV, which appears to be hot 
for the bolometric luminosity 
$1.1\times 10^{44}h^{-2}_{50}$\ergps, when compared to the 
temperature-luminosity correlation of galaxy clusters.
The irregular X-ray morphology and evidence for a 
non-uniform temperature distribution suggest that the 
system is undergoing a merger of substructures.
Since the sizes and luminosities of the individual clumps are 
consistent with those of galaxy groups,
Zw 1718.1--0108 is interpreted as an on-going merger of galaxy groups
in a dark matter halo forming a cluster of galaxies and 
thus is in a transition phase of cluster formation.
\end{abstract}


\section{Introduction}

Various surveys of clusters of galaxies have been carried out 
using X-ray techniques, since they provide an effective, unbiased
search for hot intracluster medium (ICM) bound in a cluster
potential.
The largest X-ray samples of clusters to date have been compiled from the 
ROSAT All Sky Survey (RASS) data (e.g., The Brightest Cluster Sample or BCS,
Ebeling et al 1998).
However, those surveys are usually limited to the high Galactic
latitude sky ($|b|>20^{\circ}$) to avoid confusion from Galactic 
extended sources and high Galactic extinction.
A number of clusters of galaxies located at low Galactic latitude
have therefore been missed.

We have detected a bright, extended X-ray source which is probably 
one of those missing X-ray clusters at low Galactic latitude.
The X-ray image obtained with ASCA shows extended X-ray emission 
with highly disturbed morphology ($\sim 30$ arcmin along the major axis) 
and total flux of $\sim 2\times 10^{-11}$\ergpspsqcm.
The X-ray source has been detected by earlier X-ray survey missions
such as Uhuru (4U1716--01, Forman et al 1978) and HEAO A1 
(1H1718--010, Wood et al 1984).  Since the X-ray detectors 
of those satellites had a limited imaging capability with a collimated field
of view, the extended X-ray emission was not noticed and
the radio galaxy 3C353 has been suggested as a possible optical
counterpart.
Although the cluster is listed in the Zwicky catalogue (Zw 1718.1--0108)
and the X-ray source was detected in the RASS (1RXS J172048.0--010914),
it was not selected for the 
BCS because of the Galactic latitude $19.5^{\circ}$,
just outside the survey region.

Zw 1718.1--0108 was described as ``a near, medium compact cluster'' by
Zwicky et al (1960). There are, however, few bright galaxies seen in 
the optical image as a result of the large Galactic extinction 
at the low Galactic latitude.
We carried out an ASCA observation aiming to study 3C353 and
detected hard X-ray emission from the radio galaxy as well as the
cluster emission.
We present the highly disturbed X-ray morphology 
of the cluster obtained from
ASCA and the ROSAT HRI and the X-ray spectral properties as 
evidence for a cluster merger and then discuss a possible interpretation 
on the cluster evolution.
The presence of an obscured active nucleus in 3C353 is also revealed 
through the X-ray spectrum obtained with ASCA.
No redshift measurements have been available for the member galaxies 
apart from 3C353 ($z=0.030$). In this paper, we present 
a new optical CCD image of the central region of the cluster
and the first redshift measurements for the member galaxies.

\section{Observations and data reduction}

Zw 1718.1--0108 and 3C353 were observed with ASCA and the ROSAT HRI.
A summary of the ASCA and ROSAT observations are given in Table 1.
In addition to a point-like X-ray source at the position of 
the radio galaxy, 
a bright, extended X-ray source is detected at the South-East of 
3C353 with the ASCA Gas Imaging Spectrometer (GIS; G2 and G3). 
A likely source of the extended X-ray emission is the Zwicky
cluster 1718.1--0108 of which 3C353 is a member.
Since the primary target of the observations was 3C353,
the Solid state Imaging Spectrometer (SIS; S0 and S1) of ASCA 
was operating using the standard 1 CCD chip with a $11\times 11$ arcmin
field of view which covers the radio galaxy and only a small fraction
(north part) of the cluster emission of Zw 1718.1--0108.

For the ASCA data, standard calibration 
(used for the Revision 2 processing)
and data reduction techniques were employed,
using FTOOLS (version 4.1) provided by the ASCA Guest Observer Facility
at Goddard Space Flight Center.
The GIS data were mainly used to investigate the extended X-ray emission 
from the cluster. The SIS data are used for investigating
3C353, since it provides a better spatial resolution than the GIS,
which helps to separate 3C353 from the diffuse cluster emission.


\begin{table*}
\begin{center}
\caption{The ASCA and ROSAT observations. Note that the count rates given here
contain the contribution
from 3C353. No correction for vignetting has been made.}
\begin{tabular}{lcccccc}
Satellite & Detector & Operation mode & Band & Date & Exposure & Count rate \\
          &          &                &      &      & ks & \cps \\[5pt]
ASCA &    SIS (S0/S1) & 1CCD/Faint & 0.5--10 keV &  1996 Sep 14--15  &  40.8 &
---/--- \\
     &    GIS (G2/G3) & PH    &     0.7--10 keV & & 40.1 & 0.25/0.22 \\[5pt] 
ROSAT &   HRI & &         0.1--2.4 keV & 1997 Aug 27--29 & 17.2 & 0.23 \\
\end{tabular}
\end{center}
\end{table*}

The ROSAT High Resolution Imager (Pfeffermann et al 1987) 
gave a 0.1--2.4 keV image
at the spatial resolution of about 5 arcsec.
With 17.2 ks of net exposure, a total of 5271 counts were detected.
The raw HRI image has been smoothed with the adaptive kernel
method, ASMOOTH (Ebeling, White \& Rangarajan 1999),
using a gaussian kernel and a characteristic smoothing threshold,
above the local background, of $2\sigma$.
Three X-ray clumps were detected. However, we note that  
the short HRI exposure is not appropriate to study 
the diffuse, low surface brightness X-ray emission of the cluster.

\section{Zw 1718.1--0108}
\subsection{X-ray images}


\begin{figure}
FIGURE 1.
\caption{The ASCA GIS full-band image overlaid by the ROSAT HRI
contours. Both the images were adaptively smoothed to show structures
significant at $3\sigma$ and $2\sigma$ above the local background
in the GIS and HRI images, 
respectively. 
The positions of three optically-bright galaxies are marked by plus symbols.
A bright point-like X-ray source at the North-West edge of the cluster 
emission is the radio galaxy 3C353.
Three X-ray components (N, C, and S) are indicated.}
\end{figure}



\begin{figure}
FIGURE 2a and 2b.
\caption{a) Upper panel: the ASCA GIS soft-band (0.7--2 keV) image. 
The position of 3C353 is marked by a plus symbol. The West plume (Wp) 
and North-West plume (NWp) are indicated; and b)
Lower panel: the ASCA GIS hard-band image. A strong point-like source is 
detected at the position of 3C353. Both the images have been adaptively
smoothed to show structures significant at $3\sigma$ above the local 
background. The 10 contour levels are drawn linearly between the 
minimum and the maximum of the smoothed images.}
\end{figure}


\begin{figure}
\centerline{\psfig{figure=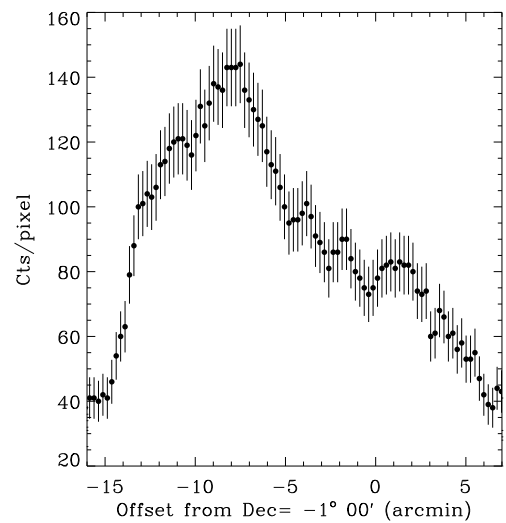,width=0.42\textwidth,angle=0}}
\caption{Projected X-ray profiles obtained from the 0.7--2 keV 
GIS image. The projection is from South to North.}
\end{figure}


\begin{table}
\begin{center}
\caption{X-ray component orientation. See Fig. 1 and 2 for naming 
conventions of X-ray components. All position angles are 
measured north through east to the system major axis.}
\begin{tabular}{ccccc}
Components & P.A. & & Component & P.A. \\[5pt]
NWp--N & 138\deg & & NWp & 140\deg \\
Wp--C & 99\deg & & Wp & 95\deg \\
C--N & 3\deg & & N & 173\deg \\
S--C & 16\deg & & C+S & 16\deg \\
\end{tabular}
\end{center}
\end{table}

The full-band GIS image superposed by the ROSAT HRI contours 
is shown in Fig. 1.
The angular scale is $\simeq 0.05$ Mpc arcmin$^{-1}$ at the redshift of 0.03
(\H0 = 50 \kmpspMpc).
The X-ray source is extended roughly in the North-South direction; 
3C353 is also detected at the NW edge of the cluster emission.
Asymmetric, slightly twisted X-ray morphology is evident in the GIS image.
Three X-ray peaks (N, C, and S from north to south) 
are resolved in the HRI image.
Faint plumes extending to the NorthWest (NWp) and West (Wp) are seen 
in the GIS image.
The NW plume is also seen at a faint level in the HRI image.
The orientation of these X-ray components are given in Table 2.

The two GIS images in the 0.7--2 keV and 5.5--10 keV bands are 
shown in Fig. 2.
3C353 is the brightest source in the hard-band image but 
faint in the soft-band image, indicating a very hard spectrum
probably due to heavy obscuration in the active nucleus (see Section 4.2).
We note that the hard energy band is similar to that (5--10 keV) 
used in the hard X-ray serendipitous source survey with BeppoSAX 
(Fiore et al 1999). 3C353 is therefore a typical source which would be selected
only by such a hard X-ray survey.
A similar hard X-ray source in a cluster is the Seyfert 2 galaxy 
NGC4388 in Virgo cluster (e.g., Iwasawa et al 1997).

The projected X-ray profile along the South-North direction,
obtained from the 0.7--2 keV GIS image, is shown in
Fig. 3. The X-ray profile is fairly flat with three peaks. 
There is no evidence for the sharply peaked X-ray emission seen in cooling
flow clusters.

The pointing accuracy of the GIS image has been checked using 
3C353 as a reference
which is clearly detected as a point-like source in the hard band image.
The displacement between the optical and GIS positions of 3C353 is 
only $\approx 5$ arcsec, well within the mean pointing uncertainty 
of 20 arcsec (Gotthelf \& Ishibashi 1997). 
A possible soft X-ray counterpart of 3C353 is barely detected in the 
HRI image, which we have used for the astromety of the HRI image.

\subsection{Optical image and redshifts of the member galaxies}


\begin{figure*}
FIGURE 4.  
\caption{The optical CCD image ($6\times 13$ arcmin$^2$) centred
on the X-ray peak of the cluster taken by the UH 2.2m telescope with
R band filter. This image is made of a mosaic of two images of overlapping
fields. Three large galaxies for which redshifts have been
measured are labeled. The results of the redshift measurements 
are listed in Table 4.
The overlaid contours show X-ray emission imaged with the ASCA GIS. Adjacent
contours show count levels that differ by factor of 1.4. The lowest visible contour is at 0.01 count arcsec$^{-1}$. All the features are significant
at $\geq 3\sigma$.}
\end{figure*}


\begin{table}
\begin{center}
\caption{Optical observations. $^a$Images of two adjacent fields were taken 
and the final image presented in Fig. 4 is a mosaic of the two images. 
$^b$This exposure time is for each field in each band. $^c$The galaxies
labeled as g1, g2 and g3 in Fig. 4 were observed.}
\begin{tabular}{ll}
\multicolumn{2}{c}{Imaging$^a$}\\[5pt]
Date & 1999 June 16 \\
Telescope & UH 2.2-m \\
Filters & R, B (Kron-Cousins system) \\
Instrument & CCD: Tektronix $2048\times 2048$\\
Pixel size & 0.22\sec\ at f/10 \\ 
Field of view & 7.5\min $\times $ 7.5\min \\
Seeing & 0.7\sec \\
Exposure time$^b$ & $3\times 120$ s \\[10pt]
\multicolumn{2}{c}{Spectroscopy$^c$}\\[5pt]
Date & 1998 October 9 (g2) \\
& 1999 July 2 (g1 and g3) \\
Telescope & UH 2.2-m \\
Instrument & Wide Field Grism Spectrograph \\
 & (low dispersion) \\
Spectral range & 4000--9000\AA \\
Resolution & $\sim 3.5$ \AA\thinspace pixel$^{-1}$ \\
Exposure time & $3\times 120$ s (g2) \\
& $3\times 300$ s (g1 and g3) \\
\end{tabular}
\end{center}
\end{table}

Follow-up optical imaging and spectroscopy of this cluster region were
carried out with the University of Hawaii 2.2-m telescope. 
The details of the observations are given in Table 3.

Fig. 4 shows the mosaic of optical R band CCD image of $6\times 13$ arcmin$^2$
field centred on the X-ray emission peak. The contours show the 
X-ray brightness obtained from the GIS image.
The redshifts of three bright galaxies, g1, g2 and g3, labeled 
in Fig. 4 were measured. The optical spectra of the three galaxies
are dominated by stellar light and their redshifts are determined with
absorption lines of NaD, MgIb (and also Balmer absorption lines in g3).
The results of the redshift measurements are shown in Table 4.
They are all consistent with $z\simeq 0.028$.
The redshift of 3C353, which is located near the North X-ray clump, is 
$z = 0.0304$, about 540 km s$^{-1}$ different in velocity 
from the brightest galaxies near the X-ray peak. From 
the observed X-ray luminosity and/or gas temperature (see Section 3.3),
we estimate a velocity dispersion of galaxies of $\sim 700$ \kmps
(cf. Fig.~1 in White, Jones \& Forman 1997). It is therefore
plausible that 3C353 and the three galaxies lie in the same system.
The extended X-ray emission detected is not a result of projection
of X-ray sources at different redshifts but physically related X-ray
emission as a whole.

The three galaxies of which we have measured redshifts
are all very red, probably due to large Galactic extinction.
The column density deduced from the HI map by Dickey \& Lockman (1990) is
\nH $\simeq 1.0\times 10^{21}$\psqcm. 
Assuming the standard gas-to-dust ratio in the Galaxy, it corresponds to
$A_{\rm V}\sim 0.5$.
An inspection of the DIRBE and IRAS 100\micron\ maps of this region
suggests the visual extinction to be even larger ($A_{\rm V}\geq 1$ mag,
J. Mulchaey, priv. comm.). Perhaps this is a dusty direction in our Galaxy.
The appearance of the cluster in Fig. 4 is of a relatively
poor cluster with a core dominated by two equally luminous ellipticals 
(g2 and g3). The number of fainter galaxies in this
field is entirely consistent with the number found in  
same size fields of other low luminosity ROSAT-selected clusters
(Edge et al., in prep.) with 8--12 galaxies in the range 2 magnitudes
fainter than the third brightest galaxy. Therefore this cluster
is unremarkable in the optical.


\begin{table}
\begin{center}
\caption{Positions and redshifts of the bright galaxies in the central
part of Zw 1718.1--0108. The galaxies, g1, g2, and g3 are denoted in 
the optical image in Fig. 4. The redshifts are of Heliocentric. 
References are 
1: De Vaucouleurs et al 1991; 2: This work.}
\begin{tabular}{ccccc}
Galaxy &  R.A.$_{\rm J2000}$ & Dec.$_{\rm J2000}$ & {\it z} & Ref \\[5pt]
3C353 &  17h20m28.1s & --00$^{\circ}$58\min 46\sec & $0.0304\pm 0.0002$ & 1\\
g1 & $17^{\rm h}20^{\rm m}40.8^{\rm s}$ & --01$^{\circ}$08\min 58\sec & 
$0.0282\pm 0.0003$ & 2 \\
g2 & $17^{\rm h}20^{\rm m}40.9^{\rm s}$ & --01$^{\circ}$11\min 57\sec &
$0.0286\pm 0.0005$ & 2 \\ 
g3 & $17^{\rm h}20^{\rm m}47.8^{\rm s}$ & --01$^{\circ}$06\min 55\sec &
$0.0286\pm 0.0006$ & 2 \\ 
\end{tabular}
\end{center}
\end{table}

\subsection{X-ray spectrum}

The spectral analysis was performed using XSPEC (version 10.0).
The MEKAL model (the original MEKA code, described by Kaastra 1992, 
with modified Fe-L line emissivity by Liedahl et al 1995)
for the optically thin, collisional ionization equilibrium 
plasma is used to fit the observed cluster emission 
with solar abundances taken from Anders \& Grevesse (1989).
The response matrices of the GIS detectors are of Version 4.0.
Since systematic errors in the low energy response of the GIS 
are suspected, we discarded the data below 0.9 keV for the present
spectral analysis.

Galactic absorption of \nH $=1.0\times 10^{21}$\psqcm,
taken from the HI survey by Dickey \& Lockman (1990), is assumed for 
all the fits presented
below unless stated otherwise, 
as no evidence for excess absorption has been found 
($\Delta N_{\rm H}\leq 0.7\times 10^{21}$\psqcm, the 90 per cent upper limit).

The GIS spectrum of the total cluster emission was 
extracted, excluding the region of 3C353, which has a very hard
spectrum.
Fitting the two GIS detectors jointly gives a temperature of $4.33\pm 0.24$ keV
and metallicity of $0.23\pm 0.10$\Zs. 
It should be noted that the radio galaxy 3C353 would affect the temperature 
measurement if it is not excluded from the photon-collection region for
the spectrum (a temperature of $\sim 5$ keV would be obtained due to 
the hard X-ray spectrum of the radio galaxy).
Even though it is unlikely that there is excess absorption above the
value derived from the HI observation, 
when the 90 per cent upper limit value of \nH 
($1.7\times 10^{21}$\psqcm) is used,
the temperature drops to $kT = 3.94\pm 0.21$ keV.
This trade-off between
temperature and absorption has little effect on 
the estimate of the bolometric luminosity given below.

The estimated observed fluxes for the whole cluster emission
are $7.0\times 10^{-12}$ \ergpspsqcm\
in the 0.5--2 keV band and $1.2\times 10^{-11}$\ergpspsqcm\ in
the 2--10 keV band. 
The bolometric luminosity corrected for the Galactic absorption is
$1.1\times 10^{44}$\ergps, using \H0\ = 50 \kmpspMpc\ and 
\q0\ = 0.5 (the 2--10 keV luminosity is $4.7\times 10^{43}$\ergps).

\subsection{Spatial variations in temperature and metallicity}


\begin{figure}
\centerline{\psfig{figure=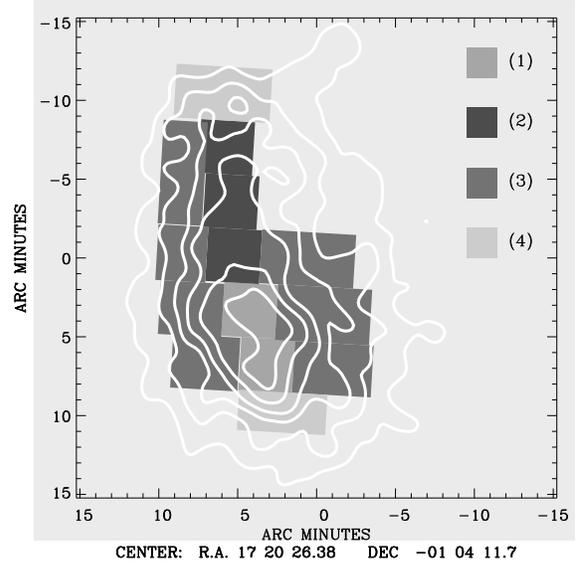,width=0.45\textwidth,angle=0}}
\caption{The ASCA GIS 0.7--2 keV map and the four regions for 
which temperature
and metallicity are investigated with the GIS spectra. Results of spectral 
fits are shown in Table 4. Note that, there is marginal
evidence for a high temperature ($kT\sim 5.4$ keV) in the middle part of
region-2, where is between the two X-ray clumps, N and C.}
\end{figure}


\begin{figure}
\centerline{\psfig{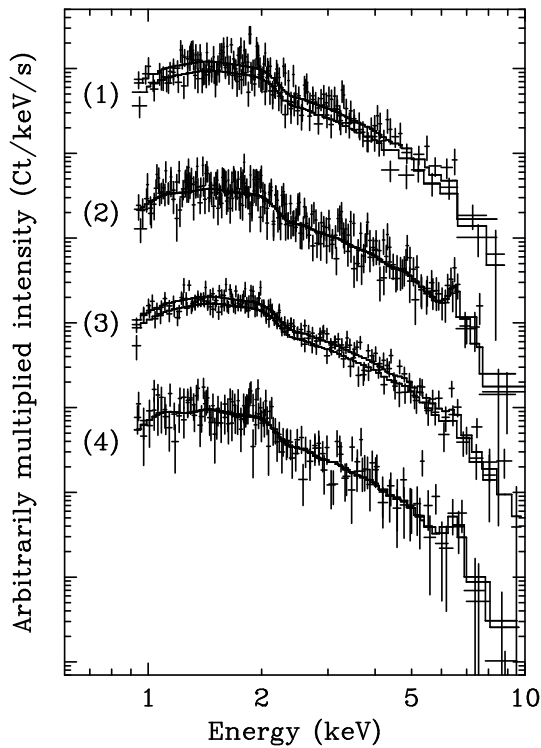}}
\caption{The GIS spectra from the four regions in Fig. 5 with the best-fit
model (solid line histograms). Intensities are arbitrarily multiplied for
clarity. Note the Fe K$\alpha$ line features.}
\end{figure}

The ASCA GIS data were investigated to search for spatial
variations in temperature and metallicity. 
A spatially resolved spectral study with the GIS data is 
severely limited by the broad PSF resulting from 
a combination of the XRT PSF which has 
a broad wing with half power diameter (HPD) of 3 arcmin 
(FWHM$\sim 20$ arcsec, see Serlemitsos et al 1995) and 
the GIS response of FWHM$\sim $1 arcmin at 2 keV. Therefore
the results presented here indicate only the qualitative trend.

The GIS image was divided into four regions as indicated in
Fig. 5, from which temperature and metallicity are derived
by fitting the 0.9--10 keV GIS spectra (Fig. 6).
The region near 3C353 was excluded from this study to avoid 
contamination by the hard X-ray source in the radio galaxy.
The spectral fits were performed using the same model as used for
the total spectrum. Results are shown in Table 5.

There is marginal evidence for a cooler temperature in
the south, the bright part (region-1), and both North and South ends of the 
cluster emission. A more detailed study provides a hint of 
a high temperature ($kT = 5.4^{+1.4}_{-0.9}$ keV) in the middle
part of region-2 [(2-m) in Table 5] which corresponds to
the intermediate region between the two X-ray clumps, N and C.

There is also some evidence for metallicity variation.
We note that the iron abundance is a primary driver of the
metallicity measurement here, since the iron K line is the
strongest line emission at the given temperature ($kT\sim 4$ keV).
The North core (region-2) and the North and South ends of the 
cluster emission (region-4) are marginally high in metallicity
($\sim $0.5--0.6 \Zs) while the south core (region-1) and
the outer part (region-3) show low metallicity ($\sim 0.15$\Zs). 

The faint NW plume (NWp), which is not indicated in Fig. 5, shows 
a sign of a much higher temperature [$kT\sim 9 (>3.8$ at 90 per cent 
confidence level) keV] or a hard spectrum.
This result is treated with caution, since the plume is
considerably faint ($\sim 0.006$ \cps) hence the spectral shape is 
sensitive to background subtraction.
No AGN-like objects can be found in NED, although it is not surprising 
for a poorly surveyed region at low Galactic latitude. 
The elongation of the X-ray image is not
consistent with the distortion of a point-like image at a large off-axis 
angle.
There are instead a few 
galaxies in the DSS image, suggesting another possible galaxy
cluster.


\begin{table}
\begin{center}
\caption{Results of spectral fits to the ASCA GIS data taken from the
whole cluster region and the four regions indicated in Fig. 5.
MEKAL is used for the spectral model with Galactic absorption of
\nH $= 1\times 10^{21}$\psqcm. The G2 and G3 data
were fitted jointly and quoted errors are
of 90 per cent confidence region for one parameter of interest.
Observed G2 count rate in the 0.9--10 keV band, temperature,
metal abundance relative to the Solar value of Anders \& Grevesse (1989),
and $\chi^2$ value with degrees of freedom are shown.
$^{\ast}$The middle region of the region-2.}
\begin{tabular}{ccccc}
Region & Count rate & $kT$ & $Z$ & $\chi^2$/dof \\
& $10^{-2}$\cps & keV & \Zs & \\[5pt]
Total & 22 & $4.33^{+0.25}_{-0.24}$ & $0.23^{+0.10}_{-0.10}$ & 562.4/573 \\[5pt]
(1) & 4.0 & $3.90^{+0.42}_{-0.36}$ & $0.16^{+0.22}_{-0.16}$ & 208.4/211 \\
(2) & 3.9 & $4.33^{+0.49}_{-0.40}$ & $0.52^{+0.29}_{-0.25}$ & 272.1/243 \\
(3) & 9.9 & $4.31^{+0.35}_{-0.32}$ & $0.15^{+0.14}_{-0.13}$ & 487.2/479 \\
(4) & 1.9 & $3.41^{+0.56}_{-0.45}$ & $0.65^{+0.62}_{-0.45}$ & 128.8/145 \\[5pt]
(2-m)$^\ast$ & 1.4 & $5.40^{+1.28}_{-0.90}$ & $0.77^{+0.65}_{-0.49}$ & 98.9/96 \\
\end{tabular}

\end{center}
\end{table}

\section{3C353}

\subsection{Powerful radio galaxy in a cluster}


\begin{figure}
\centerline{\psfig{figure=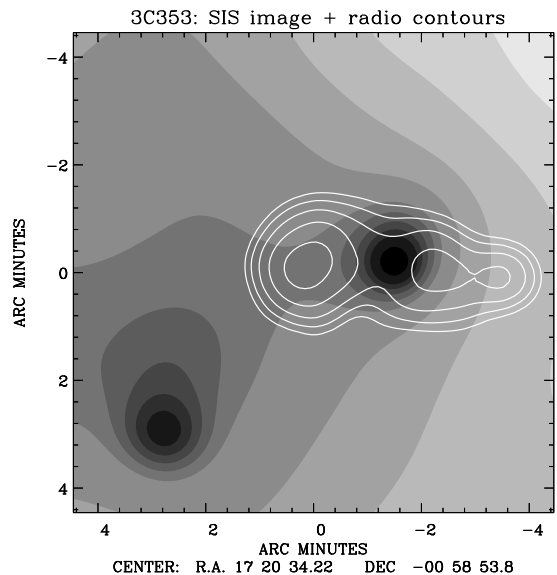,width=0.45\textwidth,angle=0}}
\caption{The radio map of 3C353 taken during the NVSS survey (contours) 
overlaid on the SIS full band image that is adaptively smoothed to 
emphasize structures significant at $5\sigma$ level. 
The bright point-like source coincides with the nucleus
of the radio galaxy. The NVSS radio image was taken by VLA at 20 cm at
45 arcsec resolution (Condon et al 1998). The 
contour levels are 0.5, 1, 2, 4 and 8 Jy per beam. 
The radio jet beaming towards 
the cluster is terminated at the bright edge near the main cluster emission.}
\end{figure}

3C353 is a powerful (log $P_{\rm 1.4GHz}\simeq 26.3$ W Hz$^{-1}$) 
FRII source with a classical double-lobe.
The radio source resides in a 
giant elliptical galaxy. 3C353 is located at the edge of the cluster X-ray 
emission (Fig. 1 and Fig. 2). The optical spectrum of nonstellar light is 
characterized by weak, low-ionization emission lines 
(Baum et al 1988; Tadhunter et al 1993).

Fig. 7 shows the radio image of 3C353 taken in the NRAO/VLA Sky Survey 
(NVSS, Condon et al 1998)
in contour superposed on the SIS full band image.
The radio image was taken by the VLA at 20 cm at resolution of 45 arcsec.
In this low resolution image, the radio source appears to be triple
while much higher resolution images such that in Morganti, Killeen \& Tadhunter
(1993) and Swain, Bridle \& Baum (1998) reveal more details 
in the radio structure as well as the weak radio core at the nucleus.
A point-like X-ray source coincides with the nucleus of the radio galaxy
which is located at the centre between the radio lobes.
The brighter jet going towards the cluster emission terminates at 
a hot spot and 
the surface brightness of the jet-side lobe steeply declines 
in front of the bright cluster emission.
The cluster medium may be acting as a working surface of the radio jet,
or it is merely an orientation effect of the jet.

\subsection{ASCA spectrum}


\begin{figure}
\centerline{\psfig{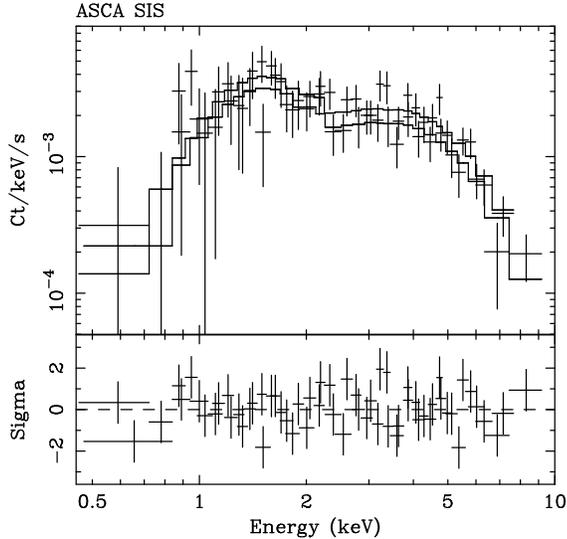}}
\caption{The ASCA SIS small-aperture spectrum of the radio galaxy 3C353.
Fitting with a power-law yields $\Gamma = 0.6\pm 0.24$ and \nH $=
4.3^{+4.2}_{-2.8}\times 10^{21}$\psqcm, demonstrating a very
hard spectrum. Note that there is a contribution from the cluster emission
in the low energy band ($\sim 60$ per cent in the 0.5--2 keV band). 
The intrinsic spectrum of 3C353 shows strong absorption of \nH
$\sim 5\times 10^{22}$\psqcm\ when the photon-index of 1.7 is assumed.}
\end{figure}

As suggested from the image analysis, the X-ray spectrum of 3C353 
is very hard. 
A likely explanation is an absorbed X-ray source of an active nucleus
in the radio galaxy.
To minimize the contamination from the cluster emission,
we use the SIS data for a spectral study as the SIS provides better
spatial resolution, and 
the data were collected from a small region with a radius
of 1.5 arcmin on the two SIS detectors centred on the hard X-ray source.
Half of the total photons from 3C353 should be contained 
in the region, when the point spread function of the ASCA XRT for 
a point source is assumed. The background data were taken from a region 
where the cluster emission is weak. The obtained spectrum should therefore 
contain some cluster emission in the soft X-ray band.

The SIS spectrum can be fitted with a very flat power-law of 
$\Gamma = 0.60\pm 0.24$ (Fig. 8). 
However, a realistic picture of the spectrum is
a sum of the cluster emission and an absorbed power-law from an AGN.
Since soft X-ray emission from the radio galaxy is very weak, as
illustrated by the HRI image and the soft band GIS image,
the energy band below 2 keV in the spectrum is dominated by the 
diffuse cluster emission.
Fitting with a model consisting of a thermal emission spectrum 
(MEKAL) with $kT=4.3$ keV and metallicity of 0.23\Zs\ and 
an absorbed power-law with a photon-index of 1.7 gives 
an absorption column density \nH $ = 5\times 10^{22}$\psqcm.
The contribution from the cluster emission (as modelled by the
MEKAL component) is found to be
about 60 per cent in the 0.5--2 keV energy range.
The absorption-corrected 2--10 keV luminosity of the X-ray
source is $8\times 10^{42}$\ergps.

An ASCA spectral study of a sample of radio galaxies including
3C353 will be reported elsewhere (see also Sambruna,
Eracleous \& Mushotzky 1999). 
Compared with the other `emission-line selected' radio galaxies,
3C353 appears to be underluminous in X-rays for its radio power.

\section{discussion}


\begin{figure}
\centerline{\psfig{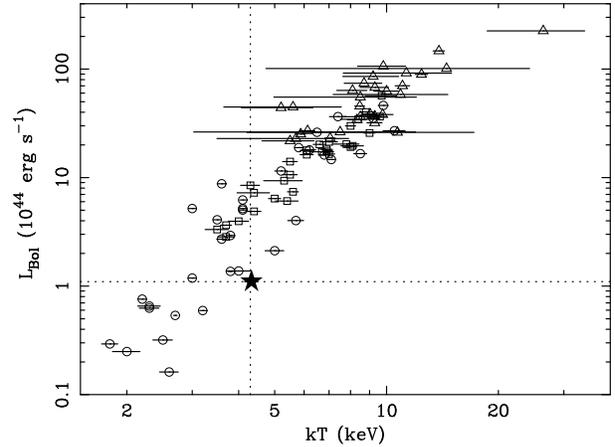}}
\caption{Plot of temperatures and bolometric luminosities of 
clusters and groups of galaxies taken from literatures; 
Triangles: Allen \& Fabian (1998); Squares: Markevich (1998);
Circles: Fukazawa et al. (1998 and priv. comm.); and Filled star: Zw1718.1--0108. 
Note that all the measurements were made with ASCA and the temperatures
have been derived taking the effect of cooling flows into account.
The luminosities are calculated with $H_0 = 50$ \kmpspMpc. The two
dotted lines are drawn at the temperature and bolometric luminosity
of Zw1718.1--0108.}
\end{figure}

We have shown that the extended emission around the powerful radio galaxy
3C353 is a cluster with highly disturbed morphology of the X-ray emission
which might affect the radio jets of 3C353. The large radio power of 3C353
could be the result of the dense environment, as 
speculated for Cyg A (Barthel \& Arnaud 1996).

This cluster, Zw 1718.1--0108, has a high temperature 
(the averaged-temperature is
$kT\simeq 4.3$ keV) for its low bolometric luminosity, $1.1 \times 
10^{44} h_{50}^{-2}$ erg s$^{-1}$ which is comparable to that
of galaxy groups rather than clusters.

To demonstrate this, we have compiled ASCA measurements of temperatures 
and bolometric luminosities of galaxy clusters and groups
from the recent literature that takes into account contamination
from cooling flows (Fig. 9, Allen \& Fabian 1998, Markevitch 1998,
Fukazawa et al. 1998 and Y. Fukazawa, priv. comm.).

Zw 1718.1--0108 lies marginally within the scatter of the correlation
which is large in the relevant luminosity range, relative to 
the scatter in the higher luminosity range.
However, Zw 1718.1--0108 is certainly one of the lowest luminosity 
clusters at the given temperature $kT\simeq $4--5 keV.

Three X-ray clumps are resolved in the X-ray images, each of which 
has a sub-Mpc size and luminosity similar to galaxy groups.
Zw 1718.1--0108 therefore appears to be a system of clustered groups,
which will collapse together to form a larger scale cluster of galaxies
as expected from the bottom-up structure formation predicted by 
standard scenario of hierarchical clustering (e.g. Tormen 1998).
The hot ICM of the system can be explained as a result of heating 
by interaction between the small clumps.   
The disturbed X-ray morphology and the unusually high temperature 
suggest that the system may not be virialized, unlike ordinary
relaxed clusters.  
The non-uniform pattern of the temperature distribution, particularly
the high temperature region between the N and C clumps, indicates 
the heating of the ICM by shock induced by interaction between the
clumps.
The low surface brightness nature of this cluster suggests that 
merging should also play a significant role to reduce bolometric luminosity
by destroying a cooling flow at the central region of the cluster.

A dramatic cluster-cluster merger is seen in Abell 754
($z=0.054$) for which a detailed numerical simulation
can reproduce the effects of the interaction between two clumps
with masses differing by a factor of about 2
(Henriksen \& Markevitch 1996; Roettiger, Stone \& Mushotzky 1998).
Zw\thinspace 1718.1--0108 is also an on-going merger but of
smaller-scale structures (galaxy-group size) in the plane-of-the-sky.
This system may be an example of transition phase of cluster evolution from
galaxy groups to a galaxy cluster. 
In particular, Doe et al. (1995) pointed out the role played by
the Zwicky clusters as ``incubators'' of compact, evolving groups:
in their sample of 5 poor clusters with physical characteristics
similar to Zw\thinspace 1718.1--0108, four of these are 
embedded within larger Zwicky clusters.


\begin{figure}
\centerline{\psfig{figure=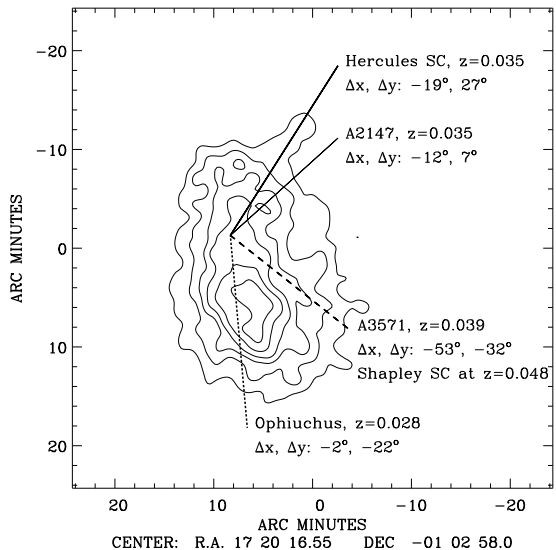,width=0.45\textwidth,angle=0}}
\caption{The contours of the 0.7--2 keV ASCA GIS image and the directions
of major clusters and superclusters at $z$ = 0.03--0.05.}
\end{figure}

Another interesting possibility is that this X-ray structure is part
of a dark matter filament, since the resolved three X-ray clumps are
linearly connected. 
The power spectrum of galaxy clustering shows 
major overdense regions to occur on a scale of 200 
$h_{50}^{-1}$ Mpc (e.g. Lin et al. 1996), 
which is $\sim 67^{\circ}$ at the redshift of 0.03. 
There are a few clusters and superclusters in the redshift range of 
0.03--0.05 which could be alined with Zw\thinspace 1718.1--0108. 
Fig. 10 shows the GIS image and directions to the Hercules supercluster
($z=0.035$), Abell 2147 ($z=0.035$), Abell 3571 ($z=0.039$), the Shapley
supercluster ($z=0.048$) and the Ophiuchus cluster ($z=0.028$). 
The Ophiuchus cluster is probably the nearest one located 22$^{\circ}$
($\sim 66$ Mpc) away to the south at the same redshift.
Although it is not clear whether they are linked to Zw 1718.1--0108,
the two nearest clusters, the Ophiuchus cluster and Abell 2147, are
moderately aligned with the elongation of Zw 1718.1--0108, which 
provides circumstantial evidence that a filament of large-scale structure
may exist.

There are several indications in favour of a low density universe
($\Omega_m\sim 0.3$, e.g., Eke et al 1998; Efstathiou et al 1999; 
Ettori \& Fabian 1999).
Most nearby clusters in such a low density universe 
are expected to be dynamically relaxed and smooth
in appearance (e.g., Richstone, Loeb \& Turner 1992; 
Buote 1998; but cf. Roettiger et al. 1998 on the role of long 
relaxation times for merger remnants). 
There are many nearby clusters, e.g., Abell 1367, Abell 2197, Abell 3627,
the Virgo cluster, and the Hercules cluster, 
that have a complex morphology with
substructures which are, however, small in size relative to 
the spherical main body.
Zw 1718.1--0108 appears to show more disturbed morphology 
than that in those nearby clusters,
and may be a rare example in the nearby universe.

Finally, it is worth pointing out that X-ray observations are an
efficient way of finding clusters of galaxies at low Galactic
latitude. Zw 1718.1--0108 is a prime example of such a cluster.
A search for clusters at $|b|<20^{\circ}$ has been conducted
using the RASS data, and has found a number of other X-ray clusters
(Ebeling, Mullis \& Tully 1999).

\section*{Acknowledgements}

We thank all the member of the ASCA team for the operation of the
satellite and maintenance of the software, John Mulchaey for
information on his unpublished CCD optical image and useful comments,
and Yasushi Fukazawa for providing the ASCA results on the cluster
sample in his thesis. This research has made use of data obtained
through the High Energy Astrophysics Science Archive Research Center
Online Service, provided by the NASA's Goddard Space Flight Center.
The NASA/IPAC Extragalactic Database (NED) is operated by the Jet
Propulsion Laboratory, California Institute of Technology, under
contract with the National Aeronautics and Space Administration.
Royal Society (ACE,ACF,SE) and PPARC (KI) are thanked for support.

\end{document}